 \documentclass[12pt,preprint]{aastex}
 
\newcommand{\masy}{${\rm mas \, yr^{-1}}$}
\newcommand{\kms}{${\rm km \, s^{-1}}$}
\newcommand{\grad}{\hbox{$^\circ$}}
\newcommand{\hehyper}{HE\,0437$-$5439}

\shorttitle{An unbound hyper-velocity B-type star}
\shortauthors{Edelmann et al.}
 
\begin{document}
 
 
\title{HE~0437$-$5439 -- an unbound hyper-velocity main-sequence B-type
star\altaffilmark{*}} 
 

\author{H.~Edelmann}
\affil{Dr. Remeis-Sternwarte, Astronomisches Institut der 
Universit\"at~Erlangen-N\"urnberg,
              Sternwartstrasse~7, D-96049~Bamberg, Germany}
\email{edelmann@sternwarte.uni-erlangen.de}
\author{R.~Napiwotzki}
\affil{Centre for Astrophysics Research, University of Hertfordshire,
College Lane, Hatfield AL10 9AB, UK}
\email{rn@star.herts.ac.uk}
%
\author{U.~Heber}
\affil{Dr. Remeis-Sternwarte, Astronomisches Institut der 
Universit\"at~Erlangen-N\"urnberg,
              Sternwartstrasse~7, D-96049~Bamberg, Germany}
\email{heber@sternwarte.uni-erlangen.de}
\author{N.~Christlieb and D.~Reimers}
\affil{Hamburger Sternwarte, Universit\"at Hamburg, Gojenbergsweg 112, 
D-21029 Hamburg, Germany}
\email{st4b321@hs.uni-hamburg.de}
\email{dreimers@hs.uni-hamburg.de}
%
%
 
 
 
\altaffiltext{*}{Based on observations collected at the European Southern
Observatory, La Silla and Paranal, Chile (Proposal No. 68.D-0192 and 70.D-0334).}
 
 
\begin{abstract}
We report the discovery of a 16$^{\rm th}$ magnitude star,
\object{\hehyper}, with a heliocentric radial velocity of $+723\pm3$~\kms. 
A quantitative spectral analysis of high-resolution optical spectra obtained
with the VLT and the UVES spectrograph shows that \hehyper\ is a main
sequence B-type star with $T_{\rm eff}$=20\,350~K, $\log(g)=3.77$, solar 
within a factor of a few 
helium abundance and metal content, rotating at $v\sin(i)=54$~\kms. Using
appropriate evolutionary tracks we derive a mass of $8~M_{\sun}$ and a
corresponding distance of 61~kpc. 
Its galactic
rest frame velocity is at least 563~\kms, almost twice the local Galactic escape
velocity, indicating that the star is unbound to the Galaxy.
Numerical kinematical experiments
are carried out to constrain its
place of birth. It has been suggested that such
hyper-velocity
stars can be formed by the tidal disruption of a binary through 
interaction with
the super-massive black hole at the Galactic center (GC).
\hehyper\ needs about 100~Myrs to travel from the GC to its present
position, much longer than its main sequence lifetime of 25~Myrs.
%
This can only be reconciled if \hehyper\ is a blue straggler star. 
In this case, the predicted proper motion is so small that it can 
only be measured by future space missions.  
Since the star is much closer to the Large Magellanic Cloud 
(LMC, 18~kpc) than to the GC, it can
reach its position from the center of the LMC. 
The proper motion
predicted in this case is about 2~\masy (relative to the LMC), large enough
to be measurable with conventional
techniques from the ground. 
The LMC origin could also be tested by a high-precision
abundance analysis.  
\end{abstract}
 
 

 
\section{Introduction}
Main sequence B-type stars located far away from the galactic 
plane are a rare,
albeit known phenomenon. They are believed to be run-away stars ejected
from the galactic plane shortly after their formation 
(see e.g. \citealt{tob87,ram01}). Up to
recently none of the 
known run-away B-type stars were found to have a velocity exceeding 
the Galactic escape velocity.
 
\citet{brow05} discovered a faint late B-type star,
SDSS~J090745.0$+$024507, 
with a heliocentric radial velocity of $853\pm12$~\kms\ (galactic rest-frame
velocity of 709~\kms). This so-called
hyper-velocity  star (HVS) is
unbound to the galaxy and \citet{brow05} conclude
that it was ejected from the Galactic center (GC) because its radial velocity
vector points 173.8\grad\ from the latter. Photometric investigations showed
it to be a slowly pulsating B-type main sequence star \citep{fuen05}.
 
\citet{hill88} predicted that velocities as high as
4\,000~\kms\ can be gained by the disruption of a binary through
interaction with the massive black hole in the galactic center \citep{scho02}.
\citet{yutr03} considered two additional processes that eject
hyper-velocity stars
in addition to the tidal breakup of binary stars by the central black hole, 
i.e. close encounters of two single stars and three-body interactions 
between a
star and a binary black hole. While the ejection rate by 
close encounters of two single stars is found to be
negligible, as much as $\sim 10^{-5}$~HVS/yr
could be ejected by the tidal breakup scenario and even
$\sim 10^{-4}$~HVS/yr    
if the Galactic center hosts a binary black hole.  
 
We report the discovery of a second hyper-velocity star,
\hehyper\ ($\alpha_{2000}=4^{\rm h}38^{\rm m}12.\!\!^{\rm s}8$, 
$\delta_{2000}=-54\grad 33^\prime 12^{\prime\prime}$), much brighter 
(16\fm2) than the HVS SDSS~J090745.0+024507.
The star was found during a spectroscopic follow-up
campaign of subluminous B (sdB) star
candidates from the Hamburg/ESO survey (see e.g. \citealt{chr01}).
 
Low resolution
spectra were obtained on 2001 November $20^{\rm th}$ at ESO, La Silla,
with the Danish 1.5~m
telescope and the DFOSC spectrograph. 
The spectrum of
\hehyper\ is very similar to a normal main sequence B-type star.
Most remarkable is its very large radial
velocity of $v_{\rm rad}=+700\pm50$~\kms.
To improve this measurement and check for radial velocity
variations we observed
the star twice again on 2002 November $14^{\rm th}$ (05:00~UT and 08:37~UT)
at the ESO Paranal observatory
using the VLT UT2 8~m telescope (Kueyen)
equipped with the UVES high-resolution echelle spectrograph 
covering the 
spectral range from 3300 to 6600\AA\ at a resolution of 0.2\AA\ and a S/N of 
5.
Radial velocities were measured with the FITSB2 program \citep{napi04}.
The spectra yield the same very high radial velocity of
$723 \pm 3$~\kms\ (721 and 726~\kms, respectively), consistent with 
that from the
low-resolution spectrum. To our knowledge it is the second largest radial
velocity ever measured for a faint blue star at high galactic latitudes.
Therefore \hehyper\ qualifies as a hyper-velocity star.
 
Since the star lies at high galactic latitude
($l_{II}=263.04$\grad, $b_{II}=-40.88$\grad) one has to consider the
possibility that it may not be a hot massive
star but a low mass evolved star that somehow mimics a main-sequence
B-type star \citep{tob87}.
Fortunately, the high resolution spectra allow us to carry out a detailed
quantitative
spectroscopic analysis, i.e. to derive temperature and gravity, elemental
abundances, and the projected rotational velocity of \hehyper\
(Section 2) in order to prove or disprove its nature as a young massive star.
Its mass and distance is then derived from
evolutionary tracks (Section 3) and the time of flight from the plane is
estimated from galactic
orbit calculations and compared to its evolutionary life time to search for
the likely place of birth (Section \ref{birth}).
\section{Spectral analysis}
\begin{figure}
\epsscale{0.5}
\plotone{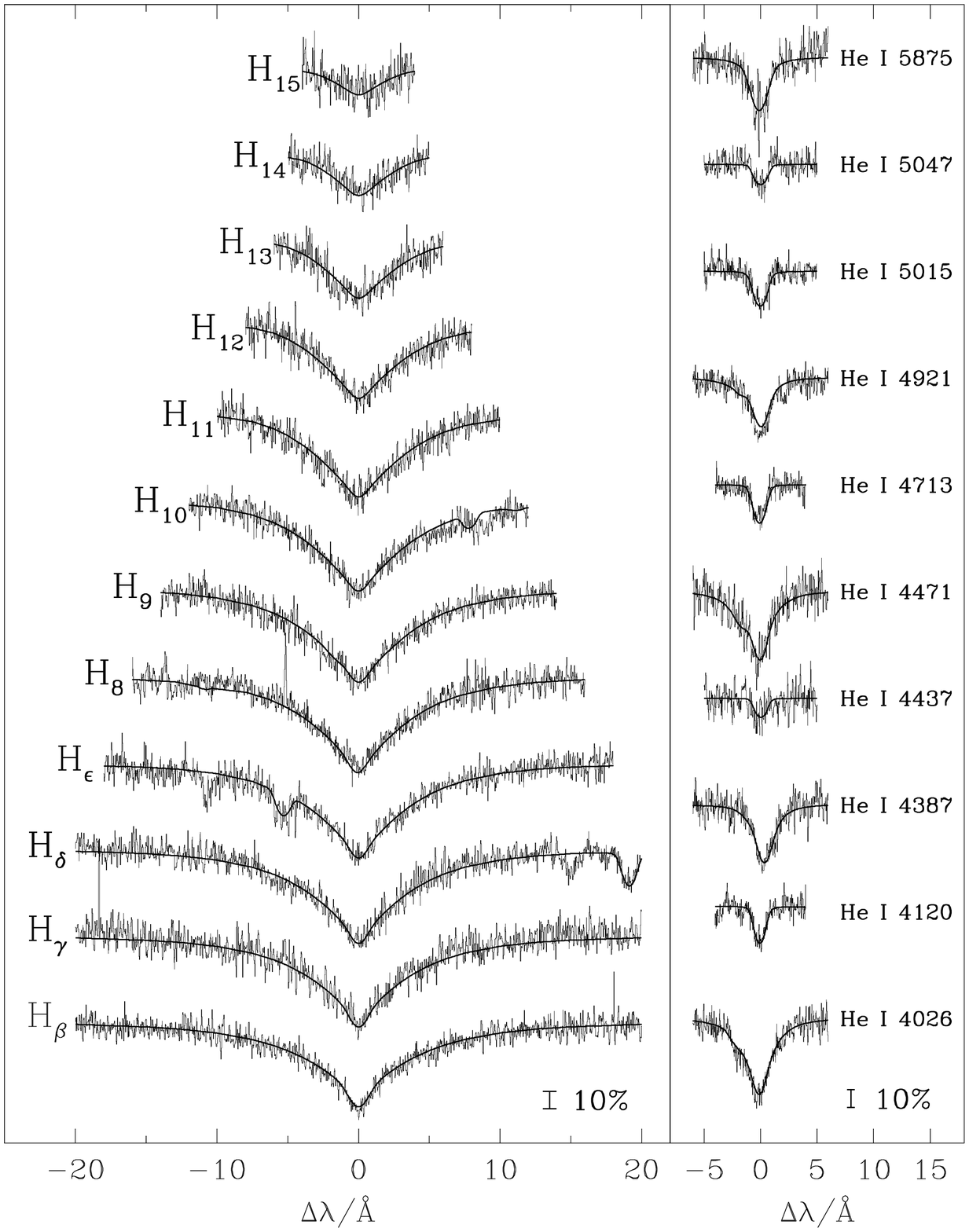}
\caption{LTE fit (thick lines) for \hehyper\ for the Co-added UVES 
spectrum (thin lines).}
\label{f1}
\end{figure}

The stellar atmospheric parameters (effective temperature $T_{\rm eff}$,
surface gravity $\log(g)$, and
photospheric helium abundance $n_{\rm He}/n_{\rm H}$)
were determined by matching synthetic line profiles
calculated from model
atmospheres to all Balmer (mainly ${\rm H}_{\beta}$ up to ${\rm H}_{15}$) and
He\,{\sc i} line profiles present in the observed spectra.
A grid of metal-line blanketed LTE model atmospheres
\citep{heb00} was used.
The models are plane parallel and chemically homogeneous
and consist of hydrogen, helium, and metals (solar abundances).
The matching procedure uses a $\chi^2$ fit
technique described by \citet{napi99}
to determine all three atmospheric parameters simultaneously.
Beforehand all spectra were normalized and
the model spectra were folded with the instrumental profile
(Gaussians with appropriate width).
The fit result is shown in Fig. \ref{f1}.

From the co-added high-resolution spectra we determined the atmospheric 
parameters to be
$T_{\rm eff}=20\,354\pm 116$~K, $\log(g)=3.77\pm 
0.02$~(cgs), and
$\log(n_{\rm He}/n_{\rm H})=-0.94\pm 0.02$. The errors are statistical ones.
The error budget, however, is dominated by systematic errors.
From our previous experiences with the analysis of similar UVES spectra
\citep{lisk05} we adopt 
$\Delta T_{\rm eff}=360$~K, and $\Delta \log(g)=0.05$~(cgs). 
%
Taken at face value
the helium abundance would be slightly above solar.
In order to verify it, we fitted the spectrum again but keeping the He 
abundance at the
solar value. The quality of the fit is as good as for the simultaneous fit
of all three parameters. Therefore, we conclude that the star's helium
abundance is solar to within 0.06~dex. Spectra of much better S/N would be
needed to narrow down the error range.
 
The rotational velocity is determined by a $\chi^2$
fit to the same lines, for which the atmospheric parameters are kept fixed.
Synthetic spectra were folded with rotational profiles (limb darkening
coefficient 0.4) for various values of $v\sin(i)$.
$v\sin(i)=54\pm4$~\kms\ ($3 \sigma$ error) results.

\begin{figure}
\epsscale{1.0}
\plotone{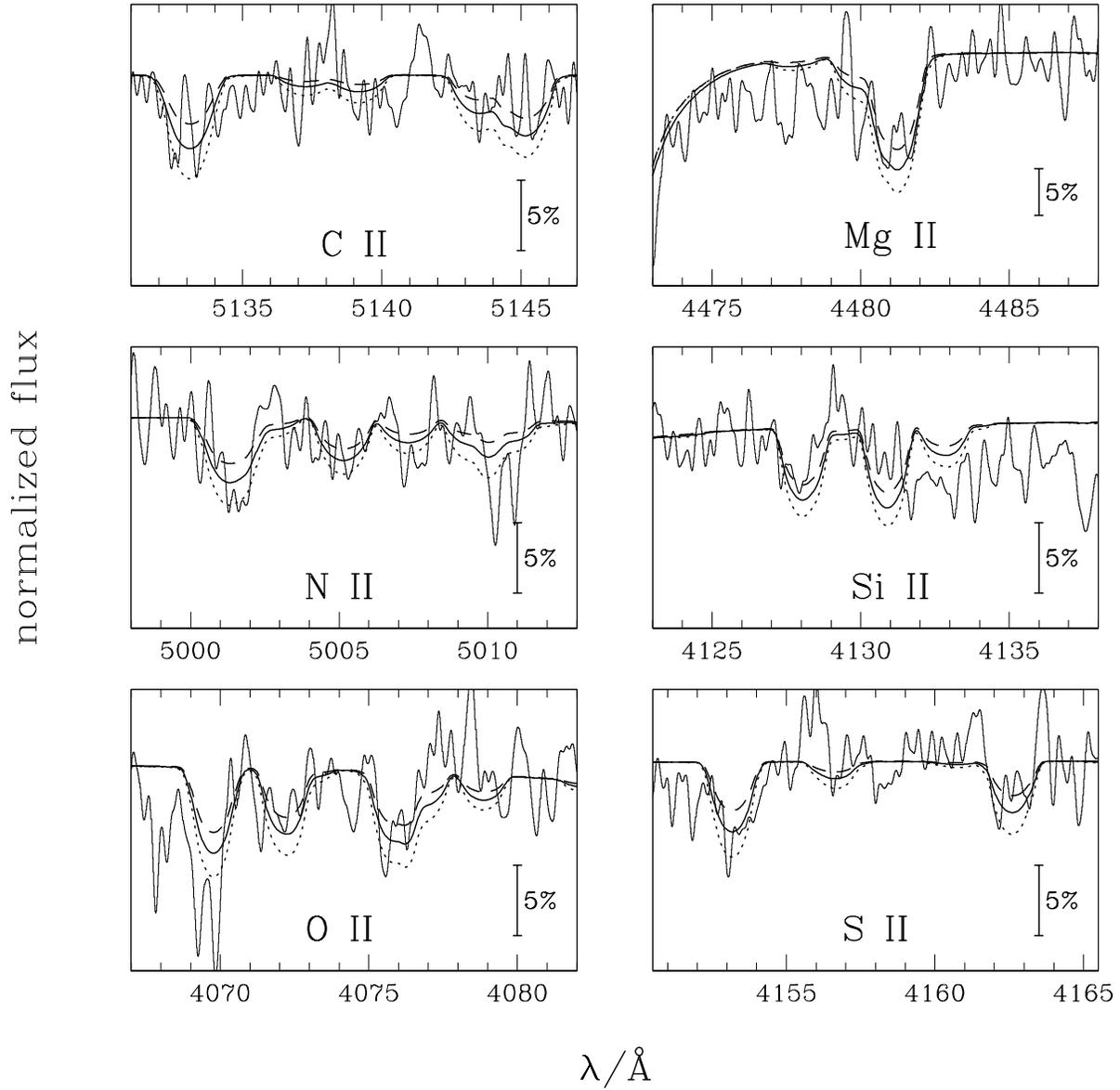}
\caption{Co-added UVES spectrum compared to synthetic spectra
with solar metal abundance (solid lines),
half the solar metal abundance (dashed lines), 
and twice the solar metal abundance (dotted lines).
Note that for this plot the spectra were binned to 0.4 \AA\
to achieve a S/N of 20.}
\label{f2}
\end{figure}

Because the exposure times were short (10~min.) the spectra are quite noisy.
Nevertheless metal lines (C\,{\sc ii}, N\,{\sc ii}, O\,{\sc ii}, 
Mg\,{\sc ii}, Si\,{\sc ii},
and S\,{\sc ii}) can  marginally be measured in the
UVES spectra. However, due to the
rotational line broadening a quantitative abundance analysis is rendered
difficult. A rough abundance estimate, however, can be
obtained by comparing synthetic spectra to the observation for solar
metallicity as well as for twice and half solar values.
The microturbulent velocity was assumed to be zero.
A solar metallicity spectrum appears to be consistent with the data
(see Fig. \ref{f2}).
 
In summary, the results of the quantitative spectral analysis strongly
suggest that
\hehyper\ is a young massive star since its effective temperature and
gravity is typical for main sequence B-type stars. The normal
metallicity supports this interpretation. While some low mass evolved
stars (HB or post-HB stars) mimic massive B-type stars to some extent 
the high rotational velocity and normal helium abundance of \hehyper\ rules out
an evolved star, because B-type horizontal branch stars are known to be
very slow rotators ($<$ 8\kms) as well as helium deficient 
\citep{behr03}.
Hence, there is no doubt that \hehyper\ is a main sequence B-type star.
\section{Mass, evolutionary timescale and distance}
\label{mass_time_dist}
\begin{figure}
\epsscale{1.0}
\plotone{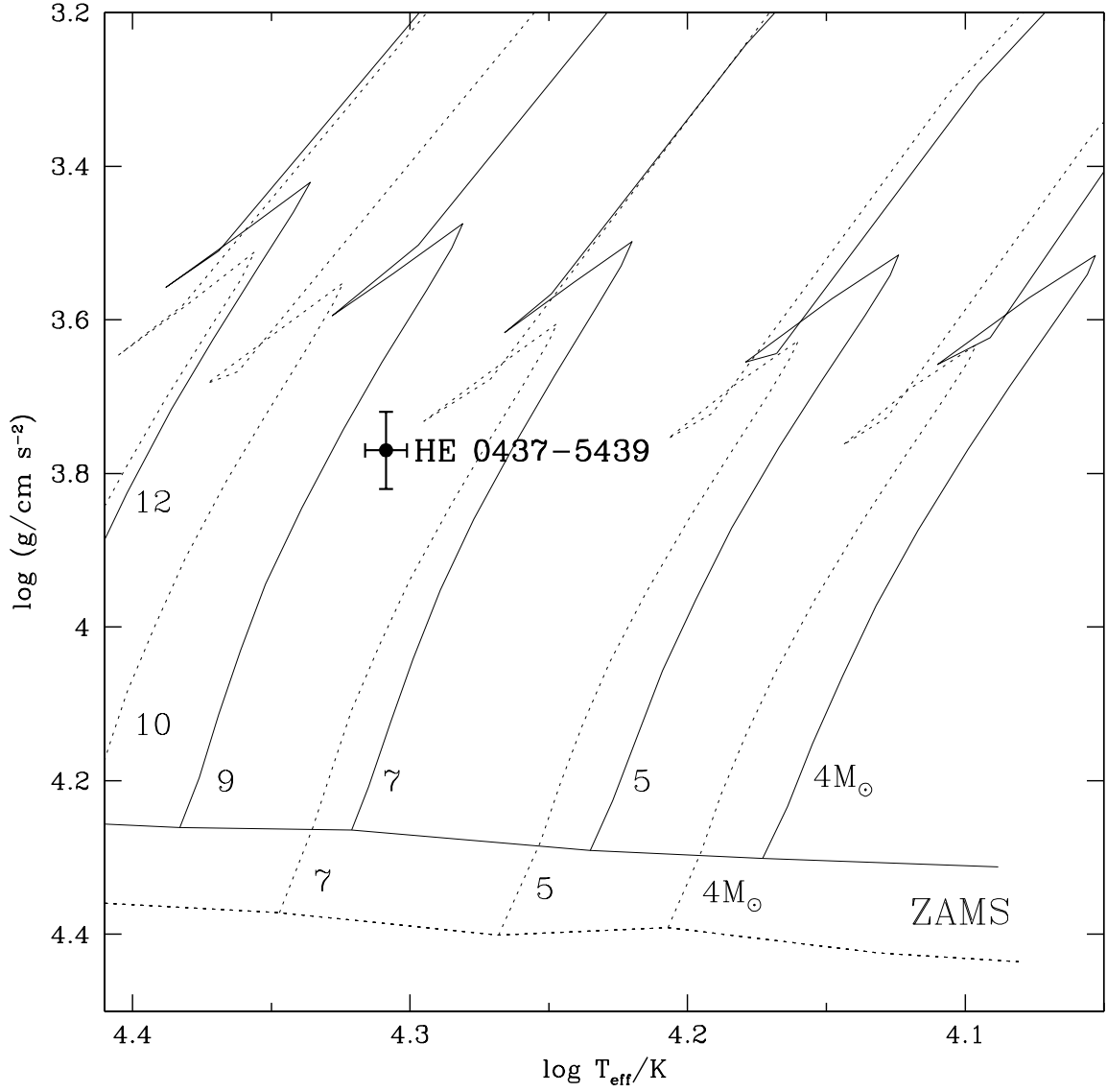}
\caption{Position of \hehyper\ in a ($T_{\rm eff}$, $\log(g)$) diagram
with evolutionary tracks for solar (\citealt{scsc92}, solid lines) 
and LMC metallicity (\citealt{scha93}, dotted lines) to
determining its mass.
\label{f3}}
\end{figure}

Having shown \hehyper\ to be a main sequence B-type star, 
we can estimate its mass by comparing its  position in the ($T_{\rm eff}$,
$\log(g)$) diagram to evolutionary tracks (see Fig. \ref{f3}). 
From solar metallicity 
models \citep{scsc92} we derive $8.4 \pm 0.5 M_{\sun}$, while a
slightly lower mass of $8.0 \pm 0.5 M_{\sun}$ results from models 
with $Z=0.008$ 
\citep{scha93}. The evolutionary time  can be
interpolated to be about 25~Myr for  solar composition and 
about 35~Myr for 
the lower metallicity (appropriate for the LMC, see Sect. 
\ref{birth}).

Using the mass, 
effective temperature, gravity and apparent magnitude
($V=16\fm2 \pm0\fm2$), we
derive the distance as described by \citet{ram01} to $d=61 \pm 12$~kpc.
 
Correcting for the solar reflex motion and to the local standard of rest, 
the Galactic velocity components can be derived from the radial velocity
($U=-55$~\kms, $V=-317$~\kms, and $W=-466$~\kms; U positive towards the 
GC and V in the direction of Galactic rotation) resulting in a 
Galactic rest-frame velocity of 563~\kms. This is a lower limit since we
assume the proper motion to be zero. 
Using the Galactic potential of \citet{alsa91} the escape velocity 
at the position of \hehyper\ is 317~\kms\ indicating that the star is 
unbound to the Galaxy.
\section{Discussion: Place of birth}
\label{birth}
 Proper motion measurements are needed to reconstruct the full space
velocity vector and trace the trajectory of \hehyper\ back to its birth
place.  The Super-COSMOS Sky Survey \citep{hamb01} 
gives $\mu_{\alpha}\cos(\delta) =
-0.6\pm 8.8$~\masy\ and $\mu_{\delta} = -6.2 \pm 6.9$~\masy,
i.e. consistent with zero and error limits larger than any
plausible value.  However, it is possible to compute the flight time
for every hypothetical origin in our Galaxy. Each starting point
corresponds to a unique set of proper motions. However, evaluation is
not straightforward, but has to be computed iteratively. Since common
wisdom has it that the most likely origin for hypervelocity stars is
the Galactic center, we started by computing the flight time to the
center of the Milky Way.
 
Calculations were performed with the program ORBIT6 developed by
\citet{odbr92}.  This numerical code
calculates the orbit of a test body in the Galactic potential.  The
complete set of cylindrical coordinates is integrated and positions
and velocities are calculated in equidistant time steps. Trial values
for the unknown proper motions were varied until the star passed
through the GC with an accuracy of better than
10\,pc. Note, that this is a formal result, which assumes a smooth
Galactic potential. However, deviations from the computed path caused
by close encounters with e.g. other stars in the central region of the
Galaxy have only negligible consequences for our considerations.
The result was $99\pm19$~Myrs with a predicted proper motion
of $\mu_{\alpha}\cos(\delta) = 0.55 $~\masy\ 
and $\mu_{\delta} = 0.09$~\masy. 
 
\begin{figure}
\epsscale{1.00}
\plotone{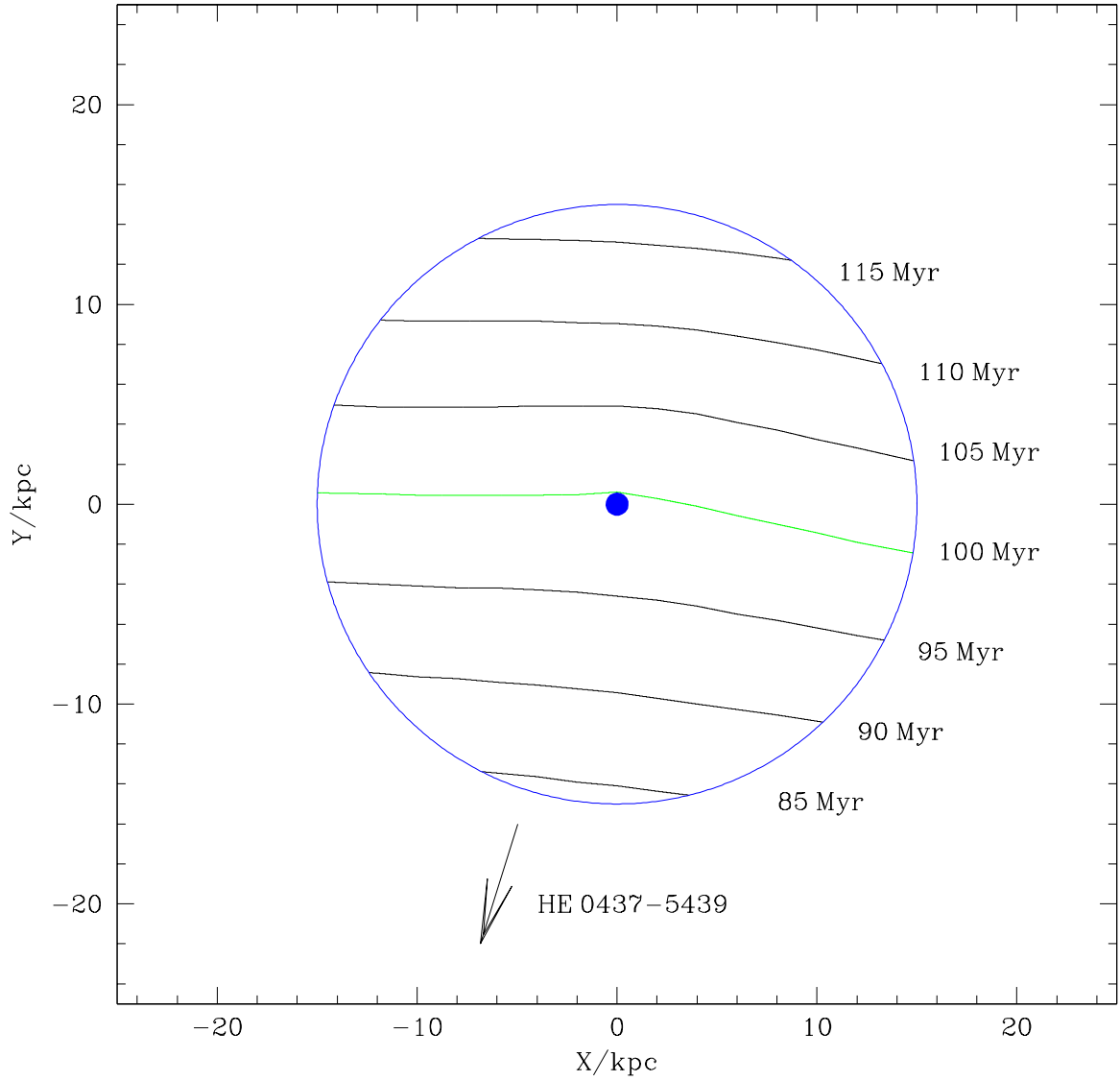}
\caption{Isochrones showing the travel time of \hehyper\ from
  hypothetical starting points in the Milky Way disk to its current
  position. X and Y are the inplane distances from the Galactic center. 
\label{f4}}
\end{figure}

Such a small proper motion can be measured
only by future space missions such as GAIA.
The time of flight is much longer than our
estimate of the main sequence age of the star (25~Myrs). This
prompted us to investigate whether an origin from another location in
the Milky Way would yield a sufficiently short time of flight. Thus we
repeated this procedure for a grid of hypothetical starting points in
the Galactic plane within a radius of 15~kpc from the Galactic center. 
The results are shown in
Fig.~\ref{f4}. Although the flight times from starting points on
the \hehyper\ side of the disk are smaller, they are always larger than
80~Myrs, i.e.\ more than three times the apparent age of \hehyper.
We performed a simple check in order to investigate how much the results
depend on the choice of a specific Galactic potential:
We repeated the procedure described above with the
numerical values of the \citet{alsa91} potential increased by
50\%. This corresponds to an upper limit for the range of escape
velocities discussed in literature (cf. \citealt{alsa91}). The
resulting flight times were slightly shorter, but never more than 5\,Myrs.
 
A possible solution of this riddle could be a blue straggler nature
of \hehyper. In this scenario a close binary consisting of two stars
with about  $4~M_\odot$ was ejected, it came into contact and merged
some time after the ejection. Since the main-sequence lifetime of 
a $4~M_\odot$ star is 165~Myrs \citep{scsc92}, 
i.e. larger than the flight time, this could resolve the
puzzling age discrepancy. In principle it would be possible to eject a
 close binary after the interaction of a hierarchical triple system
 with the central black hole of the Milky Way. However, the necessary
 kick velocity is so high, that probably a lot of fine tuning would
 be necessary to allow the binary system to survive. 
 
Theoretical calculations testing this scenario are encouraged.
We want to point out here that a
LMC origin is a feasible alternative.
 
The projected distance
of \hehyper\ to the kinematic center of the LMC
($\alpha = 5^{\mathrm{h}} 27\farcm 6$, $\delta=-69\degr 52\arcmin$; 
\citealt{vdma02}) is $16.3\degr$. This 
corresponds to 14~kpc at the distance of the LMC (50~kpc;
\citealt{free01}). Our distance estimate for \hehyper\ was $61\pm
12$~kpc which puts it $11\pm 12$~kpc behind the LMC. We
derive a total distance between \hehyper\ and the center of LMC of
$18_{-4}^{+9}$~kpc. Thus \hehyper\ is much closer to the center of the
LMC than to the disk of the Milky Way. 
 
\Citet{vdma02} determined a systematic velocity of the
LMC of 262~\kms\ from a study of a large sample of carbon stars. Thus
the relative (radial) velocity of \hehyper\ is 461~\kms.  If \hehyper\
was ejected shortly after its birth about %
35~Myrs ago from somewhere close
to the center of the LMC (see Sect. \ref{mass_time_dist}) 
a tangential velocity of 
390~\kms\ would be
necessary to explain its current position $16\degr$ away from the
center. This corresponds to a proper motion of about 2~\masy, which could
be measured by conventional methods from the ground.
 
The total ejection velocity would amount to 600~\kms\
(neglecting the gravitational potential of the LMC) close to the value 
computed for an ejection from the Milky Way disk.
We conclude that the properties of \hehyper\ could be well explained by
the ejection of a single star from the LMC about 
35~Myrs ago. It is
not necessary to invoke a blue straggler scenario, 
which requires a fair amount of fine tuning. 
\section{Conclusions and outlook}
We have shown that \hehyper\ is an early B-type hyper-velocity
star. Both, chemical abundances and a moderate rotational velocity of
\hehyper\ provide strong evidence for a main sequence nature of this
object. It is considerably more massive than the first object of this
class, SDSS~J090745.0$+$024507, which puts tight constraints on the 
flight time since \hehyper's ejection. 
Tidal disruption of a close binary by the massive black hole at the GC has
been suggested to explain the HVS. However, an origin from the center 
or the disk of
the Milky Way is at variance with its age. Although it is possible to
reconcile the age constraint via a blue straggler scenario, we showed
that an ejection from the LMC is more plausible.
 
Our LMC scenario makes two predictions: firstly, we expect that the
abundances of \hehyper\ should correspond to the LMC metallicity, which
is about half solar. Secondly,  we predict a proper motion relative to
the LMC of 2~\masy\ or higher, depending on the precise location and
time of the \hehyper\ ejection. 
Unlike the proper motion predicted for ejection from the Galactic center, 
the 
proper motion is large enough to be possibly measurable from the ground. 
 
If an origin in the LMC can be confirmed, this would allow to proof that
either the LMC contains a -- so far undetected -- very massive black hole, 
or that other mechanisms are capable of producing HVS as well.
 
\begin{acknowledgements}
We thank the staff of the ESO La Silla and Paranal observatories for their
valuable support.
We made extensive use of NASAs Astrophysics Data System Abstract Service
(ADS) and the SIMBAD database, operated at CDS, Strasbourg, France. 
R.~N. acknowledges support by a PPARC Advanced Fellowship.
\end{acknowledgements}
\clearpage
 
\clearpage
 
\clearpage
 
\end{document}